\documentclass[aps, prl, twocolumn, showpacs, nofootinbib, showkeys]{revtex4-1}

\usepackage{amssymb} \usepackage{amsmath} \usepackage{graphicx}
\usepackage{epsfig,latexsym}

\RequirePackage{xspace} \allowdisplaybreaks

\begin{document}

\def\bef{\begin{figure}}
\def\eef{\end{figure}}

\newcommand{\nl}{\nonumber\\}
\newcommand{\ans}{ansatz }
\newcommand{\be}[1]{\begin{equation}\label{#1}}
\newcommand{\beq}{\begin{equation}}
\newcommand{\ee}{\end{equation}}
\newcommand{\beqn}[1]{\begin{eqnarray}\label{#1}}
\newcommand{\eeqn}{\end{eqnarray}}
\newcommand{\bd}{\begin{displaymath}}
\newcommand{\ed}{\end{displaymath}}
\newcommand{\mat}[4]{\left(\begin{array}{cc}{#1}&{#2}\\{#3}&{#4}
\end{array}\right)}
\newcommand{\matr}[9]{\left(\begin{array}{ccc}{#1}&{#2}&{#3}\\
{#4}&{#5}&{#6}\\{#7}&{#8}&{#9}\end{array}\right)}
\newcommand{\matrr}[6]{\left(\begin{array}{cc}{#1}&{#2}\\
{#3}&{#4}\\{#5}&{#6}\end{array}\right)}
\newcommand{\cvb}[3]{#1^{#2}_{#3}}
\def\lsim{\raise0.3ex\hbox{$\;<$\kern-0.75em\raise-1.1ex
e\hbox{$\sim\;$}}}
\def\gsim{\raise0.3ex\hbox{$\;>$\kern-0.75em\raise-1.1ex
\hbox{$\sim\;$}}}
\def\abs#1{\left| #1\right|}
\def\simlt{\mathrel{\lower2.5pt\vbox{\lineskip=0pt\baselineskip=0pt
           \hbox{$<$}\hbox{$\sim$}}}}
\def\simgt{\mathrel{\lower2.5pt\vbox{\lineskip=0pt\baselineskip=0pt
           \hbox{$>$}\hbox{$\sim$}}}}
\def\unity{{\hbox{1\kern-.8mm l}}}
\newcommand{\eps}{\varepsilon}
\def\ep{\epsilon}
\def\ga{\gamma}
\def\Ga{\Gamma}
\def\om{\omega}
\def\omp{{\omega^\prime}}
\def\Om{\Omega}
\def\la{\lambda}
\def\La{\Lambda}
\def\al{\alpha}
\newcommand{\ov}{\overline}
\renewcommand{\to}{\rightarrow}
\renewcommand{\vec}[1]{\mathbf{#1}}
\newcommand{\vect}[1]{\mbox{\boldmath$#1$}}
\def\tm{{\widetilde{m}}}
\def\mcirc{{\stackrel{o}{m}}}
\newcommand{\Dm}{\Delta m}
\newcommand{\dm}{\varepsilon}
\newcommand{\tanb}{\tan\beta}
\newcommand{\nbar}{\tilde{n}}
\newcommand\PM[1]{\begin{pmatrix}#1\end{pmatrix}}
\newcommand{\up}{\uparrow}
\newcommand{\down}{\downarrow}
\def\omE{\omega_{\rm Ter}}
%
%%%%%%%%%%     mauri    %%%%%%%%%%%%%%%%%%%%%%%%%%%%%%%%%

\newcommand{\Dsusy}{{susy \hspace{-9.4pt} \slash}\;}
\newcommand{\DCP}{{CP \hspace{-7.4pt} \slash}\;}
\newcommand{\mc}{\mathcal}
\newcommand{\gr}{\mathbf}
\renewcommand{\to}{\rightarrow}
\newcommand{\gtc}{\mathfrak}
\newcommand{\wh}{\widehat}
\newcommand{\br}{\langle}
\newcommand{\kt}{\rangle}

%%%%%%%%%%%%%%%%%%%%%%%%%%%%%%%%%%%%%%%%%%%%%%%%%%%%%%%%%%

% barbara Ricci  %definizione di minore e maggiore simile
\def\lsim{\mathrel{\mathop  {\hbox{\lower0.5ex\hbox{$\sim$}
\kern-0.8em\lower-0.7ex\hbox{$<$}}}}}
\def\gsim{\mathrel{\mathop  {\hbox{\lower0.5ex\hbox{$\sim$}
\kern-0.8em\lower-0.7ex\hbox{$>$}}}}}
%%%%%%%%%%%%%%%%%%%%%%%%%%%%%%%%%%

\def\nn{\\  \nonumber}
\def\de{\partial}
\def\brf{{\mathbf f}}
\def\bbf{\bar{\bf f}}
\def\bF{{\bf F}}
\def\bbF{\bar{\bf F}}
\def\bA{{\mathbf A}}
\def\bB{{\mathbf B}}
\def\bG{{\mathbf G}}
\def\bI{{\mathbf I}}
\def\bM{{\mathbf M}}
\def\bY{{\mathbf Y}}
\def\bX{{\mathbf X}}
\def\bS{{\mathbf S}}
\def\bb{{\mathbf b}}
\def\bh{{\mathbf h}}
\def\bg{{\mathbf g}}
\def\bla{{\mathbf \la}}
\def\bmu{\mathbf m }
\def\by{{\mathbf y}}
\def\bmu{\mbox{\boldmath $\mu$} }
\def\bsig{\mbox{\boldmath $\sigma$} }
\def\bunity{{\mathbf 1}}
\def\cA{{\cal A}}
\def\cB{{\cal B}}
\def\cC{{\cal C}}
\def\cD{{\cal D}}
\def\cF{{\cal F}}
\def\cG{{\cal G}}
\def\cH{{\cal H}}
\def\cI{{\cal I}}
\def\cL{{\cal L}}
\def\cN{{\cal N}}
\def\cM{{\cal M}}
\def\cO{{\cal O}}
\def\cR{{\cal R}}
\def\cS{{\cal S}}
\def\cT{{\cal T}}
\def\eV{{\rm eV}}
%
%%%%%%%%%%%%%%%%%%%%%%%%%%%%%%%%%%%%%

%\title{Testing Dark Matter models with Radio Astronomy}
\title{Testing Dark Matter Models with Radio Telescopes in light of Gravitational Wave Astronomy}

\author{Andrea Addazi$^1$}\email{andrea.addazi@lngs.infn.it}
\author{Yi-Fu Cai$^{2,3}$}\email{yifucai@ustc.edu.cn}
\author{Antonino Marcian\`o$^1$}\email{marciano@fudan.edu.cn}
%\author{Dong-Gang Wang$^{4,5}$}\email{wdgang@strw.leidenuniv.nl}
\affiliation{$^1$ Department of Physics \& Center for Field Theory and Particle Physics, Fudan University, 200433 Shanghai, China}
\affiliation{$^2$ CAS Key Laboratory for Researches in Galaxies and Cosmology, Department of Astronomy, University of Science and Technology of China, Hefei, Anhui 230026, China}
\affiliation{$^3$ School of Astronomy and Space Science, University of Science and Technology of China, Hefei, Anhui 230026, China}
%\affiliation{$^4$ Leiden Observatory, Leiden University, 2300 RA Leiden, The Netherlands}
%\affiliation{$^5$ Lorentz Institute for Theoretical Physics, Leiden University, 2333 CA Leiden, The Netherlands}

\begin{abstract}
\noindent
In this Letter we put forward a novel phenomenological paradigm in which particle physics beyond the Standard Model may be tested by radio astronomy if they are related to a first order phase transition in the early Universe.
For this type of Dark Matter models, the first order phase transition takes place at KeV scales, and hence, induces the production of a stochastic gravitational wave background that can be detected from Pulsar timing measures.
We demonstrate this hypothetical feasibility by studying a class of Majoron Dark Matter model, which is related to a first order phase transition of the $U(1)_{L}$ or $U(1)_{B-L}$ symmetry and is consequently dubbed as {\it violent Majoron}.
This phenomenon are expected to be examined by the ongoing and forthcoming radio experiments, including FAST, SKA and IPTA.
\end{abstract}


% \keywords{Dark Matter, Lepton/Baryon number violation, Majorons, radio astronomy, gravitational waves}

\maketitle

%\section{Introduction}
%\label{s.intro}

\noindent
{\it Introduction.--}
Almost a century ago, the hypothesis of Dark Matter (DM) was proposed to meet with a variety of astronomical and cosmological observations that is invisible through electromagnetic interactions. The nature of DM remains one of the biggest mysteries in physics today. In the literature there are many theoretical proposals and experimental designs (see Refs. \cite{Jungman:1995df, Bertone:2004pz, Feng:2010gw} for comprehensive reviews). The hypothesis of the weakly interacting massive particles (WIMP) is the prevailing paradigm. Also, there are alternative models in terms of the axions and axion-like particles (ALP). Another possibilities allow the existence of the heavy hidden sector particles that only interact with ordinary matter fields through gravitation. Depending on theoretical properties of these candidates, the associated DM experiments can be divided into two classes: the direct detection, which searches for the scattering of DM particles with atomic nuclei; and the indirect detection, which searches for the particle physics productions such as the annihilation or decay effects of DM particles.

Along with the arrival of the gravitational wave (GW) astronomy, the detection method of DM deserves to be revisited since more information about DM might be revealed by testing GW signals if these candidate particles can be related to a first order phase transition (FOPT) in the early Universe.
This is because that a FOPT can generate a characteristic stochastic background of GWs. In this mechanism the Universe was initially in a state of false vacuum. Then, the tunneling toward the state of true vacuum could induce enucleation and percolation of bubbles, which expand with constant acceleration, driven by a difference of pressure between the interior true vacuum and the exterior false vacuum. Eventually, these bubbles produce the stochastic GW background in the Universe through the processes of scattering, turbulence and acoustic shock waves.

From the perspective of observations, the energy scale, or equivalently, the temperature of the FOPT is crucial for the possible detection in various astronomical instruments. It determines the characteristic frequency window of GWs. For instance, for phase transitions that occurred around $100\, {\rm GeV} \sim  1{\rm TeV}$, the GW signal is peaked within the range of $1 \sim 10\, {\rm mHz}$
\cite{Witten:1984rs, Turner:1990rc, Hogan:1986qda, Kosowsky:1991ua, Kamionkowski:1993fg, Hindmarsh:2013xza, Hindmarsh:2015qta, Delaunay:2007wb}.
This region of frequencies will be experienced by the next generation of interferometers, including LISA, U-DECIGO and BBO \cite{Caprini:2015zlo, Kudoh:2005as, Audley:2017drz}. Theoretical models associated with a FOPT with the energy scale at about  $100\, {\rm GeV} \sim 1{\rm TeV}$ were recently discussed in Refs. \cite{Schwaller:2015tja, Chala:2016ykx, Huber:2015znp, Huang:2016odd, Artymowski:2016tme, Dev:2016feu, Katz:2016adq, Addazi:2016fbj, Katz:2016adq, Baldes:2017rcu, Chao:2017vrq, Ghorbani:2017jls, Tsumura:2017knk, Huang:2017rzf, Addazi:2017gpt, Addazi:2017oge, Cheung:2017lpv, Cai:2017tmh}.
It is interesting to note that, for this type of dark FOPT arisen from DM models, the energy scale is theoretically allowed to be much lower than the above regime, and thus, the corresponding GW signals would become elusive for GW interferometers. For example, the energy spectra of GWs generated from the MeV-ish or KeV-ish FOPT are peaked within $10^{-6} \sim 10^{-3}\, {\rm mHz}$, which is out of scope of the current GW interferometers including LIGO/VIRGO, LISA, U-DECIGO, BBO and so on.

In this Letter we report a brand new paradigm that DM models with a dark FOPT occurred at low energy scale, which are in the blind zone of the GW interferometers, are able to be tested by radio telescopes. In light of early works on radio astronomy \cite{Sazhin:1978, Detweiler:1979wn, Hellings:1983fr, Foster:1990, Shannon:2015ect, Lentati:2015qwp, Liu:2015psa}, it is acknowledged that GW signals within the frequency range of $10^{-9} \sim 10^{-7}\, {\rm Hz}$ are detectable by virtue of Pulsars timing data. Using this effect we can probe the DM models that can give rise to FOPTs within the energy scale range of $10^{-1} \sim 10\,{\rm KeV}$. To demonstrate this phenomenon, we specifically consider a type of Majoron DM scenario in which a FOPT is naturally envisaged for the extremely low frequency range.

In particle physics, Majorons are a hypothetical type of pseudo-Nambu-Goldstone boson originated from a spontaneous symmetry breaking of a global $U(1)_{L}$ or $U(1)_{B-L}$ symmetry by extending the gauge group of the Standard Model (SM). It belongs to a complex scalar singlet that is coupled to a Majorana neutrino operator. After the spontaneous symmetry breaking, a Majorana mass for the neutrino is generated\footnote{Implications of the Majoron in neutron-antineutron transitions were discussed recently \cite{Berezhiani:2015afa, Addazi:2015pia, Addazi:2015ata, Addazi:2016rgo}.} \cite{Chikashige:1980ui, Gelmini:1980re, Schechter:1981cv}. As shown in \cite{Akhmedov:1992hi}, a Majoron is possible to be coupled with visible matter very weakly and its mass could be of the KeV-ish scale, and hence, it can be viewed as a good candidate of warm DM \cite{Akhmedov:1992hi, Berezinsky:1993fm}. A variety of cosmological bounds can be imposed on the Majoron models as a matter of fact that the spontaneous symmetry breaking scale of $B-L$ and $L$ shall be higher than the electroweak vacuum expectation value ({\it vev}). Compatible with pre-sphaleron baryogenesis models, the Majoron couplings to visible matter are highly constrained if the $B-L$ scale is higher than the electroweak scale \cite{Steigman:1979xp, Olive:1980wz, Olive:1989xf, Walker:1991ap, Cline:1993ht}. On the other hand, from the Majoron overproduction bounds, stringent constraints can be over-imposed on the Majoron models with a $B-L$ phase transition higher than the $10 {\rm TeV}$ scale \cite{Akhmedov:1992hi}. Intriguingly, a sub-electroweak phase transition can delicately avoid the aforementioned cosmological bounds. An open possibility left is that such a phase transition is a first order one. In this case, a FOPT within $0.1 \sim 10\, {\rm KeV}$ is achieved straightforwardly and becomes dark since it is almost invisible in colliders.

{\it The model.--} We consider an extension of the SM characterized by the gauge groups $SU_{c}(3)\times SU(2)_{L}\times U(1)_{Y}\times U(1)_{B-L}$. In this model the lepton and baryon numbers are encoded in a new $U(1)_{B-L}$ global symmetry, while a complex scalar field coupled to neutrinos and to the Higgs boson via the following potential:
\begin{equation}
%\label{fff}
 %\mathcal{L}_{M} =
 fH\bar{L}\nu_{R} +h\sigma \bar{\nu}_{R}\nu_{R}^{c} +h.c. +V(\sigma,H) ~, \nonumber
\end{equation}
with $h$ and $f$ being Yukawa matrices of the model, spontaneously breaks the $U(1)_{L}$ symmetry. The potential $V(\sigma,H)$ is responsible for the {\it vev} of the scalar singlet $\sigma$, {\it i.e.} $\langle \sigma \rangle=v_{BL}$, and  triggers the generation of a Majorana mass term $\mu\,\bar{\nu}_{R}\nu_{R}+h.c.$, where $\mu = h v_{BL}$ (see \cite{Addazi:2017oge} for details).
The scalar sector of the model is contributed by the new scalar singlet $\sigma$, containing the Majoron field in its imaginary part, and by the Higgs boson. Thus, the scalar potential can be organized as
\begin{equation}
%\label{VsH}
 V(\sigma,H) = V_{0}(\sigma,H) +V_{1}(\sigma) +V_{2}(\sigma, H) ~, \nonumber
\end{equation}
where
\begin{align}
%\label{VsH2}
 V_{0}(\sigma,H) =& \lambda_{s} (|\sigma|^{2} -\frac{v_{BL}^{2}}{2})^{2} +\lambda_{H} (|H|^{2} -\frac{v^{2}}{2})^{2} \nonumber\\
 &+\lambda_{sH} (|\sigma|^{2}-\frac{v_{BL}^{2}}{2}) (|H|^{2}-\frac{v^{2}}{2}) ~, \nonumber
\end{align}
and $V_{1,2}$ are higher order effective operators that are expected to trigger the FOPT.

We analyze below two possibilities: {\it i)} the FOPT is triggered by five-dimensional (5-d) effective operators that softly break the $U(1)_{B-L}$ symmetry; {\it ii)} 5-d operators are suppressed, and instead, the FOPT is triggered by the six-dimensional (6-d) interactions.
For the first case, the 5-d operators can be expressed as:
\begin{align}
%\label{Vos}
 V_{1}^{(5)}(\sigma) = & \frac{\lambda_{1}}{\Lambda}\sigma^{5} +\frac{\lambda_{2}}{\Lambda}\sigma^{*}\sigma^{4} +\frac{\lambda_{3}}{\Lambda}(\sigma^{*})^{2}\sigma^{3} +h.c ~, \nonumber \\
%\label{Vdhs}
 V_{2}^{(5)}(\sigma, H) = &
 \frac{\beta_{1}}{\Lambda}(H^{\dagger}H)^{2}\sigma +\frac{\beta_{2}}{\Lambda}(H^{\dagger}H)\sigma^{2}\sigma^{*} \nonumber\\ & +\frac{\beta_{3}}{\Lambda}(H^{\dagger}H)\sigma^{3}
 +h.c. ~; \nonumber
\end{align}
while, for the latter one, the 6-d operators are given by,
\begin{align}
\label{sesto}
 V_{1}^{(6)}(\sigma) = & \frac{\gamma_{1}}{\Lambda^{2}}\sigma^{6} +\frac{\gamma_{2}}{\Lambda^{2}}\sigma^{*}\sigma^{5} +\frac{\gamma_{3}}{\Lambda^{2}}(\sigma^{*})^{2}\sigma^{4} \nonumber\\
 & +\frac{\gamma_{4}}{\Lambda^{2}}(\sigma^{*})^{3}\sigma^{3} +h.c. ~, \\
\label{sesto2}
 V_{2}^{(6)}(\sigma, H) = & \frac{\delta_{1}}{\Lambda^{2}}(H^{\dagger}H)^{2}\sigma^{2} +\frac{\delta_{2}}{\Lambda^{2}}(H^{\dagger}H)^{2}\sigma^{*} \sigma \nonumber\\
 & +\frac{\delta_{3}}{\Lambda^{2}}(H^{\dagger}H)\sigma^{3}\sigma^{*} +\frac{\delta_{4}}{\Lambda^{2}}(H^{\dagger}H)(\sigma\sigma^{*})^{2} \nonumber\\
 & +\frac{\delta_{5}}{\Lambda^{2}}(H^{\dagger}H)\sigma^{4} +h.c. ~.
\end{align}
In principle, the energy scale of new physics entering non-perturbative operators may be different from each other. For convenience, we parameterize their differences in the couplings $\lambda_{i}$, $\beta_{i}$, $\gamma_{i}$ and $\delta_{i}$. One may also consider the case of $B-L$ preserving effective operators, which would entail all the operators introduced in Eqs. (\ref{sesto}) and (\ref{sesto2}) to be hyper-selected. As a result, only 6-d operators remain and are associated with the parameters: $\gamma_{4},\delta_{2}, \delta_{4}$.

%\section{Gravitational Waves signal from Majoron DM}
{\it GW signals from Majoron DM.--} The spontaneous symmetry breaking of the $U(1)_{B-L}$ is related to the FOPT in the early Universe, despite of a second order phase transition. This phenomenon can generate vacuum bubbles expanding at relativistically high velocity, and can in turn induce a stochastic background of gravitational radiation. GWs are produced via three main processes, i.e., bubble-bubble collisions, turbulence induced by the bubbles' expansions in the cosmic plasma, and, sound waves induced by the bubbles' running though the plasma. These three contributions are directly related to the thermally corrected effective potential. Specifically, the energy spectrum of GWs produced during the collision of two bubbles depends only on the grossest features of the collision, which are the bubble-size at collision and the false-vacuum energy. Results then carry a parametric dependence only on the temperature of the FOPT and the ratio between the rate of change of the bubble enucleation rate and the rate of expansion of the Universe. The peak frequency of GW signals produced by bubbles' collisions corresponds to \cite{Kosowsky:1991ua,Kamionkowski:1993fg}
\begin{equation}
\label{nucollision}
 f_{\rm c}\simeq 3.5 \times 10^{-4}\,\! \Big( \frac{\beta}{H_{*}} \Big) ~ \Big( \frac{\bar{T}}{10\, {\rm GeV}} \Big) ~ \Big( \frac{g_{*}(\bar{T})}{10} \Big)^{1/6} ~ {\rm mHz}\,,
\end{equation}
in which $\beta$ is related to the size of the bubble wall and is expressed in \eqref{deba}, $\bar{T}$ is the temperature at the FOPT, $g_{*}(\bar{T})$ labels the degrees of freedom that are involved. In this situation, the GW intensity is estimated as \cite{Kosowsky:1991ua,Kamionkowski:1993fg}
\begin{eqnarray}
\label{CCC}
 \Omega_{\rm c}(\nu_{\rm c}) \simeq C_{\rm c} \mathcal{E}^{2} \Big( \frac{\bar{H}}{\beta} \Big)^{2} ~ \Big( \frac{\alpha}{1+\alpha} \Big)^{2} ~ \frac{V_{B}^{3}}{0.24+V_{B}^{3}} ~ \frac{10}{g_{*}(\bar{T})}  ~,
\end{eqnarray}
with $V_{B}$ representing for the velocity of the bubble, which is linked to the size of the bubble wall $\beta$ by the relation $V_B \simeq \beta d$ with constant $d$. Different values of $V_{B}$ can yield different amount of corrections from turbulence and sonic waves. The coefficient $C_{\rm c}$ can be numerically estimated as $2.4\times 10^{-6}$. Moreover, the rest parameters are determined by the following relations:
\begin{align}
% \label{E}
 & \mathcal{E}(\bar{T})= \Big[ T\frac{dV_{\rm eff}}{dT}-V_{\rm eff}(T) \Big]|_{T=\bar{T}} ~, \nonumber \\
% \label{a}
 & \alpha = \frac{\mathcal{E}(\bar{T})}{\rho_{\rm rad}(\bar{T})} ~,~~ \rho_{\rm rad}=\frac{\pi^{2}}{30}g_{*}(T)T^{4} ~, \nonumber
\end{align}
where $\rho_{\rm rad}$ stands for the energy density of radiation and there is $\bar{T}\simeq v_{BL}$ which denotes the FOPT temperature. The later is defined by
\begin{equation} \label{deba}
 \beta = - \Big[ \frac{dS_{E}}{dt} \Big]_{t=\bar{t}}\simeq \Big[ \frac{1}{\Gamma}\frac{d\Gamma}{dt} \Big]_{t=\bar{t}} ~,
\end{equation}
in which
\begin{align}
 &S_{E}(T) \simeq \frac{S_{3}(T)}{T} ~,~~ S_{3} \equiv \int d^{3}r [ \partial_{i}\sigma^{\dagger}\partial_{i}\sigma+V_{\rm eff}(\sigma,T) ] ~, \nonumber \\
 &\Gamma_{0}(T) \sim T^{4} ~,~~ \Gamma = \Gamma_{0}(T) \exp[-S_{E}(T)] ~. \nonumber
\end{align}
Specifically, thermal corrections to the effective potential can be calculated within the approximation performed in Ref. \cite{Delaunay:2007wb}, namely,
%\begin{equation}
%\label{Veff}
 $ V_{\rm eff}(s,T)\simeq CT^{2}(\sigma^\dagger \sigma)+ V(\sigma,H) $.
%\end{equation}
Note that, for the case of 5-d effective operators,
\begin{equation}
\label{coeff}
 C \equiv C^{(5)} = \frac{1}{4} \Big( \frac{m_{\sigma}^{2}}{v_{BL}^{2}} +\frac{m_{H}^{2}}{v^{2}} +h^{2}-24K_{BL} \Big) ~,
\end{equation}
with
%\begin{equation}
%\label{coeff2}
 $ K_{BL}\equiv K_{BL}^{(5)}=(\lambda_{2}+\lambda_{3})\frac{v_{BL}}{\Lambda}+\beta_{2}\frac{v_{BL}}{\Lambda} $.
%\end{equation}
For 6-d operators, the expression of $C$ in Eq.~(\ref{coeff}) remains unchanged, but the form of $K_{BL}$ is replaced by
%\begin{equation}
%\label{coeff3}
 $ K_{BL} \equiv K_{BL}^{(6)} = \frac{1}{\Lambda^{2}} [(\delta_{2}+\delta_{3} +\gamma_{2} +\gamma_{3} +\gamma_{4}) v_{BL}^{2} +(\delta_{2} +\delta_{3}) v^{2} ] $.
%\end{equation}

\begin{figure}[t]
\centerline{ \includegraphics [height=5.cm,width=0.9\columnwidth]{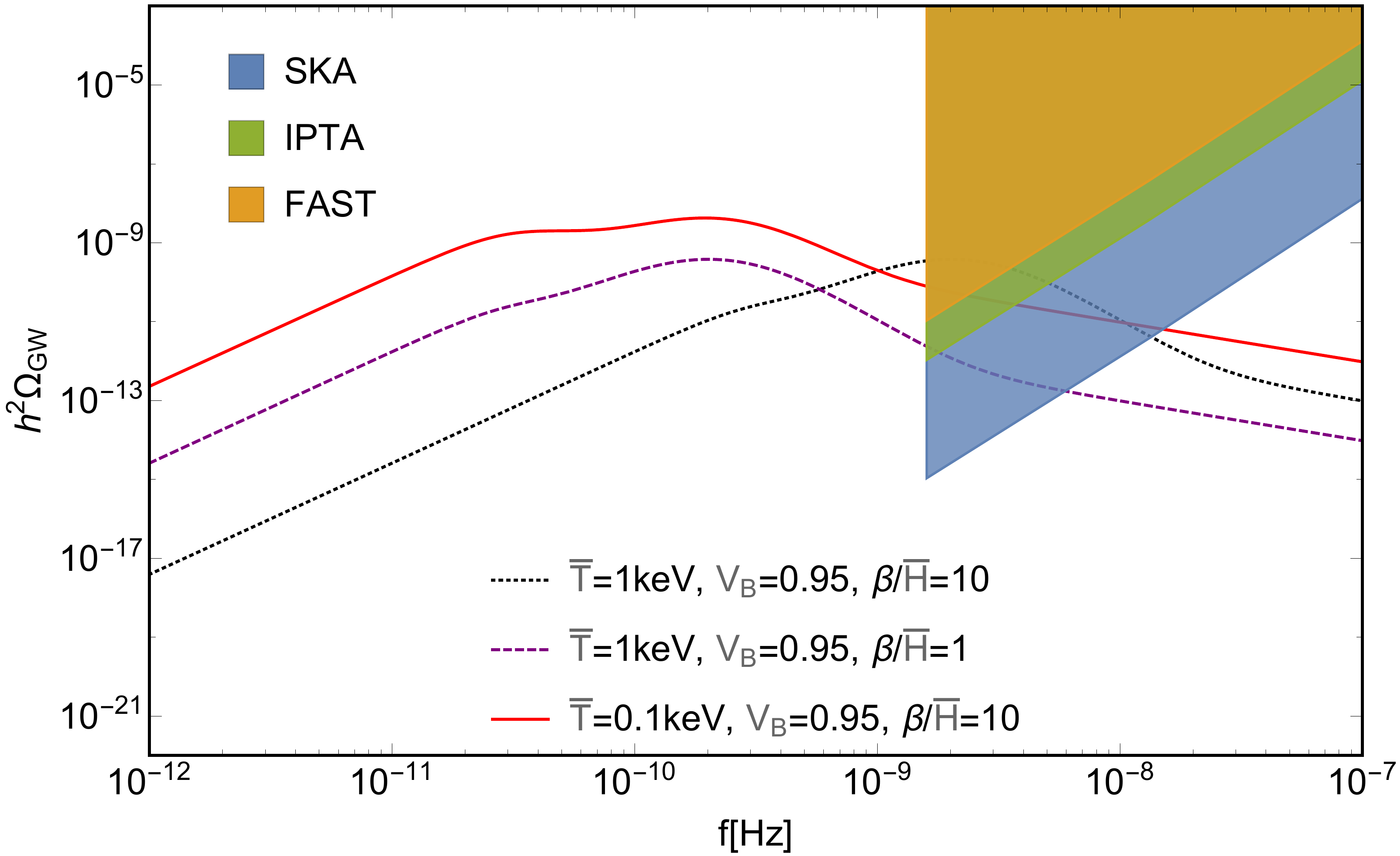}}
\caption{
Numerical results of three examples of energy spectra produced from the dark FOPTs and their comparisons with radio astronomy.
We denote the FOPT temperature with $\bar{T}$,
the Hubble parameter at $\bar{T}$ with $\bar{H}$,
the size of the bubble wall with $\beta$,
and the bubble velocity with $V_B$.
We display three representative curves labelled by different parameter choices as follows:
$( \bar{T}\simeq 1 ~ {\rm KeV}$, $V_{B}=0.95$, $\beta/\bar{H}=10 )$ in black dotted;
$( \bar{T}\simeq 1 ~ {\rm KeV}$, $V_{B}=0.95$, $\beta/\bar{H}=1 )$ in purple;
$( \bar{T}\simeq 10^{-1} ~ {\rm KeV}$, $V_{B}=0.95$, $\beta/\bar{H}=10 )$ in red.
}
\label{plot}
\end{figure}

In Fig.~\ref{plot}, we numerically simulate the GW signals produced from the FOPT arisen from the model of Majoron DM. Considering different bubble velocities, we focus on 5-d effective operators that softly break the $B-L$ global symmetry and induce a FOPT. We obtain a bound for the parameter space of Majoron as
\begin{equation}
\label{Majoron}
 (\lambda_{2} +\lambda_{3} +\beta_{2}) \frac{v_{BL}}{\Lambda} \lesssim 4 \times 10^{-2} \Big[ \frac{m_{\sigma}^{2}}{v_{BL}^{2}} +\lambda_{sH} +h^{2} \Big] ~,
\end{equation}
in which the above model parameters are provided in the previously introduced model building.

For the KeV-ish scale Majorons specified by the assumptions with $v_{BL} \simeq 1 {\rm KeV}$, $\lambda_{sH} \ll 1$, $h \simeq 1$, and $\lambda_{i=1,2,3}\simeq O(1)$ as well as $\beta_{j=2,3}\simeq O(1)$, a bound $\Lambda \gtrsim 24\, v_{BL}$ can be recovered. Nonetheless, another possibility is that the FOPT is triggered by 6-d operators instead of 5-d ones. In this case, the bound on the 6-d operators is derived to be,
\begin{equation}
\label{Majoron2}
 K_{BL}^{(6)} \lesssim 4\times 10^{-2} \Big[ \frac{m_{\sigma}^{2}}{v_{BL}^{2}} +\lambda_{sH}+h^{2} \Big] ~.
\end{equation}

The remarkable point of the FOPT triggered by 6-d operators is that $U(1)_{B-L}$ can be exactly preserved.
In this case, the only possible parameters entering in $K_{BL}^{(6)}$ are $\delta_{2}$ and $\gamma_{4}$. Since the electro-weak scale is much larger than the $B-L$ scale, the expression of $K_{BL}^{(6)}$ can be much simplied as:
%\begin{equation}
%\label{FBL}
 $K_{BL}^{(6)} \simeq \frac{\delta_{2}v^{2}}{\Lambda^{2}} $.
%\end{equation}
%
Assuming that $\delta_{2} \simeq 1$, $\lambda_{sH} \ll 1$, $h \simeq 1$ and $v_{BL} \ll v$, it turns out that $\Lambda \gtrsim 24\, v$. This implies that, since the Higgs {\it vev} insertion enters in the 6-d operators, the new physics scale $\Lambda$ shall be bounded from below, which is close to the TeV scale.

It is worth noticing that the possibility addressed in the previous analysis is easily compatible with the Majoron DM model. In particular, 5-d operators can generate a mass term for the Majoron as \cite{Akhmedov:1992hi}:
\begin{equation}
\label{Majoron44}
 m_{\chi} = \Big( \frac{\beta_{1}v}{v_{B-L}} \Big)^{1/2} {\rm keV} ~.
\end{equation}
For $v_{BL} \simeq {\rm keV}$ and $\beta_{1}$ varies from $10^{-6}$ to $O(1)$, the Majoron mass could be within the range $1{\rm keV} \sim 100\, {\rm MeV}$. For this range of mass scale, it is possible to generate sizable GW signals that are sensitive to both present and forthcoming radio experiments. 

In Fig.~\ref{plot}, three curves depict different parameter choices for GW energy spectra that arise from violent Majorons FOPTs. We compare the theoretical curves with current bounds from FAST \cite{FASTweb} and sensitivity regions that will be probed by  SKA \cite{SKAweb} and IPTA. Our results displayed in Fig.~\ref{plot} can be easily understood from Eqs.~(\ref{nucollision}, \ref{CCC}). For example, for the red curve with relativistic bubble velocities $V_{B}=0.95$, bubble nucleation ratio $\beta/H=10$ and B-L VEV scale $v_{BL}=0.1\, {\rm keV}$, we obtain a GW peak around $\nu \simeq 2\times 10^{-10}\, {\rm Hz}$ with an intensity of $\Omega_{GW} h^{2}\simeq 10^{-9}$, in agreement with semi-analytical results. Within the case of the black dotted result, the spectrum has a peak that is roughly one order of magnitude higher in frequency, while on the same GW intensity scale of the red curve. Indeed, the intensity of the two signals is controlled by the same parameters $\beta/H$, $V_{B}$ and $\alpha$, which are fixed by the form of the potential. On the other hand, the frequency is shifted by one order of magnitude, being proportional to the FOPT temperature. Similar conclusion can be reached while considering the purple curve. In the latter case, the GW intensity is one order of magnitude smaller than in the case shown by the red curve, while the peak lies in the same range. This is again in agreement with the semi-analytical estimates Eqs.~(\ref{nucollision}, \ref{CCC}).

We remark that GW spectra generated by KeV-scale FOPTs do not violate experimental bounds from CMB. The GW signals predicted from violent Majoron models are suppressed down to $\Omega h^{2}\simeq 10^{-22} \sim 10^{-23}$ around $10^{-15} ~ {\rm Hz}$, which is much smaller than the current Planck bound. Moreover, it has been studied in \cite{Berezinsky:1993fm, Rothstein:1992rh} that, for the same mass range the observationally expected amount of DM can be produced from several different mechanisms in the Universe.

%\section{Conclusions and remarks}
%\noindent
{\it Conclusions and remarks.--} We explored the paradigm of testing DM models that are related to the FOPT in the early Universe, from radio astronomy constraints. These phase transitions occur around the KeV-ish scales and can give rise to a stochastic GW background, but the associated signals are not sensitive to traditional GW interferometers.
However, it is remarkable to note that, they may be measured by the Pulsar timing data, of which the peak frequency lies in the low frequency band.
We studied a specific type of the Majoron DM models, which are related to baryon-lepton symmetry phase transitions at KeV-ish scales.

Radio instruments, such as FAST, SKA and IPTA, may observationally test the Majoron DM or impose more stringent bound on their parameter space. These signals from the Majoron model are not in contradiction with any CMB constraints. Our analysis fully demonstrate the possibility of indirectly probing DM physics by virtue of radio technology. 
In the near future, a combination of radio astronomy and particle collider physics could become crucial to extract hidden information of the dark world. Finally, in the era of multi-messenger astronomy, GW signals from Majoron DM may be compared with more indirect dark matter bounds, and therefore, may motivate further future experiments, including e-ASTROGAM \cite{DeAngelis:2017gra}.

%\section*{Acknowledgments}
%\noindent
{\it Acknowledgments.--}
We are grateful to D. Bastieri, M. Bianchi, P. Chen, J. Ellis, M. Sasaki, G. Veneziano, D.G. Wang and Y. Wan for valuable communications.
AM acknowledges the support by the Shanghai Municipality through the grant No. KBH1512299 and by Fudan University through the grant No. JJH1512105.
YFC is supported in part by the Chinese National Youth Thousand Talents Program, by the NSFC (Nos. 11722327, 11653002, 11421303, J1310021), by the CAST Young Elite Scientists Sponsorship Program (2016QNRC001), and by the Fundamental Research Funds for the Central Universities.
%DGW is supported by a de Sitter Fellowship of the Netherlands Organization for Scientific Research (NWO).
%Part of numerical computations are operated on the computer cluster LINDA in the particle cosmology group at USTC.

\end{document}